\documentclass[12pt]{article}
\usepackage{epsfig}
\usepackage{amsmath}
\usepackage{amsfonts}
\usepackage{amssymb}
\usepackage{amsthm}

\theoremstyle{plain}

\begin{document}

\begin{center}
 \textbf{A Conceptual Framework for} \\
 \textbf{Understanding Faster-Than-Light Neutrinos} \\
 Eric Sakk$^1$ and Aradhya P. Kumar$^2$ \\
$^1$Department of Computer Science \\
eric.sakk@morgan.edu \\
$^2$Department of Physics \\
Morgan State University, Baltimore, MD, 21251 \\
aradhya.kumar@morgan.edu \\
\end{center}

\vspace{.3in}

\footnotesize
\begin{center}
\textbf{Abstract}
\end{center}
\noindent Recent experiments have led to the production of neutrinos
with transit times indicating the appearance of traveling faster
than the speed of light.   In this paper, we present a conceptual
framework to understand how faster-than-light events involving
neutrinos (as indicated by time-of-flight) might occur.   We propose
that observations of this kind do not violate the special theory of
relativity; instead, they only help to provide evidence in support
of the general theory of relativity at quantum scales. Given the
relativistic effects of the neutrino on its local spacetime
environment, the measured time-of-flight at the macroscopic level is
attributable to a decrease in the effective path length traversed by
the neutrino. Specifically, along preferred directions, we show that
the Kerr metric allows for the compression of spacetime; hence, the
decreased path length hypothesis is plausible. Furthermore, when the
motion of the neutrino is along the preferred direction for
spacetime compression, the Kerr metric also predicts strong frame
dragging effects near the Planck length. In the region where strong
frame dragging occurs, we propose that the microscopic explanation
for the path length compression is due to the formation of
'micro-wormholes' near the Planck length.

\vspace{.3in}

\normalsize
\section{Introduction} \label{sec:intro}
A recent set of high energy experiments and associated measurements appear to
indicate that neutrinos can travel faster than the speed of light \cite{adam,reich}.
Such observations are reminiscent of and potentially consistent with reports of neutrinos
detected from supernova events \cite{arnett,antonioli,dasgupta,paglia}, The scientific
community as a whole is on very solid ground in its unwillingness to entertain the idea
that particles can be energized to exceed the speed of light. Clearly, either the
observations are incorrect or there exists a sensible explanation that could lead to a
deeper understanding of neutrino behavior.  In the event that the experimental results
are correct, it is sensible to seek after a more accurate description of particle behavior
that would not violate the theory of relativity.

The purpose of this work is to offer a possible resolution that
necessarily requires describing the interaction of the neutrino with
its spacetime surroundings in a manner consistent with the general
theory of relativity.  This setting is ideal for considering only
the gravitational effect of an uncharged particle possessing nonzero
angular momentum. We therefore propose that the particle behavior
can be approximated by considering effects derivable directly from
the Kerr metric; namely, path length compression and frame dragging.
Given the possibility of a spacetime compression, the measured
time-of-flight at the macroscopic level is attributable to a
decrease in the effective path length traversed due to
neutrino-spacetime interactions at high energies.

The 'faster-than-light' claim has arisen in many contexts.  A common
example comes from tunneling experiments with photons
\cite{winful,nimtz}. Quantum mechanics does not provide a complete
description detailing how the photon arises on the other side of a
barrier; however, it is clear that the decreased optical path length
leads to the macroscopic illusion of faster-than-light group
velocity. While the mechanism we present in this work is due to
gravitational effects, the principle of a decreased path length is
the same. For this work, in addition to the spacetime compression
observable from a macroscopic rest frame, we further propose that
the microscopic explanation lies in frame dragging effects also
predicted by the Kerr metric. Specifically, we hypothesize the
formation of 'micro-wormholes' to account for tunneling behavior at
microscopic scales near the Planck length.

Section \ref{sec:pathlength} describes the path length compression
effect as derived from existing experimental data. Section
\ref{sec:metric} then derives conditions where the Kerr metric
predicts spacetime compression along preferred directions of motion.
Section \ref{sec:standardmodel} then briefly investigates why other
particles within the standard model have not exhibited similar
behavior. Section \ref{sec:framedragging} continues this discussion
by analyzing frame dragging effects for charged versus uncharged
leptons. It is then pointed out that, for the Kerr metric at scales
near the Planck length, strong frame dragging occurs along the same
directions where spacetime compression is demonstrated in Section
\ref{sec:metric}. Finally, in addition to discussing various
implications of the neutrino behavior, Section \ref{sec:discussion}
raises the idea of 'micro-wormhole' formation to account for
particle tunneling behavior at microscopic scales near the Planck
length.

\section{Path Length Analysis} \label{sec:pathlength}
The traversal time $t_\nu$ of the neutrino over a length $\ell$ has recently been measured to be less than that of a photon $t_p$ \cite{adam,reich},
\begin{equation} \label{eqn:tinequality}
t_\nu = \frac{\ell}{\beta c} < \frac{\ell}{c} = t_p
\end{equation}
(where $\beta >1$), leading to the perception of faster-than-light neutrino velocity $\beta c$. Viewed in this light, it is possible to arrive at the incorrect conclusion that special relativity is being violated.  Instead, we describe a general relativistic formalism that characterizes the effect of the neutrino on its spacetime surroundings. Specifically, as measured from the reference frame of an observer at rest, we propose that the effective path length decreases to a value of $\ell' \equiv \gamma_\nu \ell$ where $\gamma_\nu < 1$ is the neutrino path length compression factor.  Furthermore, according to special relativity, the actual neutrino velocity as observed from a rest frame must be $v_\nu = \eta_\nu c$ where $\eta_\nu < 1$.  Hence, the traversal time is more accurately described as
\begin{equation} \label{eqn:tactual}
t_\nu = \frac{\ell'}{v_\nu} = \frac{\gamma_\nu \ell}{\eta_\nu c} = \frac{\ell}{(\frac{\eta_\nu}{\gamma_\nu} ) c}
\end{equation}
Comparing this result with Equation (\ref{eqn:tinequality}), imposes the condition
\begin{equation} \label{eqn:betacondition}
\beta=\frac{\eta_\nu}{\gamma_\nu}  > 1 \Rightarrow \eta_\nu > \gamma_\nu.
\end{equation}
From  \cite{adam,reich}, the assumed velocity $v_a=\beta c$ in Equation (\ref{eqn:tinequality}) can be related to the experimentally observed values:
\begin{equation}
\frac{v_a-c}{c} = \frac{\beta c-c}{c} = \beta - 1 = 10^{-5}.
\end{equation}
However, given Equation (\ref{eqn:betacondition}), we propose that the condition
\begin{equation} \label{eqn:alphagammacondition}
\frac{\eta_\nu}{\gamma_\nu} - 1 = 10^{-5}
\end{equation}
more accurately characterizes the observed neutrino effect.
Since $\eta_\nu$ is a knowable quantity (e.g. $\eta_\nu \approx .999995$), we can isolate the value of the neutrino path length compression factor $\gamma_\nu$ in order to describe the general relativistic effects on spacetime leading to the value
\begin{equation} \label{eqn:gammacondition}
\gamma_\nu = \frac{\eta_\nu}{1+10^{-5}}.
\end{equation}

As another example, consider data recorded from Supernova 1987A \cite{arnett} where $\ell \approx 168,000$ light years and neutrinos reached detectors on earth approximately 3 hours ($\approx 10^4$ sec) before the supernova was visible.  Under these circumstances, using the above formulation, the arrival time difference can be characterized as
\begin{equation} \label{eqn:chksupernova}
\Delta t=t_p - t_\nu= \frac{\ell}{c}-\frac{\ell}{\beta c}=\frac{\ell}{c}(1-\frac{1}{\beta}).
\end{equation}
Assuming $(v_a-c)/c \approx 10^{-9}$ for neutrinos ejected from the supernova, we arrive at value of $\beta=1+10^{-9}$. According to Equation (\ref{eqn:chksupernova}),
\begin{equation}
\Delta t=\frac{(1.68 \times 10^5)(9.64 \times 10^{15})}{3 \times 10^8}(1-\frac{1}{\beta}) = 5 \times 10^3 \; s
\end{equation}
which approaches the observed arrival time difference between the neutrinos and photons.

\section{Neutrino-Spacetime Interactions at High Energy} \label{sec:metric}
\subsection{Kerr Metric} \label{sec:Kerr metric}
The mean free path length for the neutrino is enormous; hence, the path length compression effect is cumulative and observable only at high energies. In addition, since the neutrino is a chargeless lepton, we hypothesize that the compression effect is likely due to gravitational interactions at extremely small scales. To this end, we propose the Kerr metric \cite{oneill,steph} as a starting point for understanding the behavior of a gravitational mass possessing angular momentum interacting with its spacetime surroundings.  Specifically, the Kerr metric for a mass $M$ with angular momentum $J$ can be written using spherical coordinates $(r,\theta,\phi)$ as:
\begin{equation} \label{eqn:Kerrmetric}
c^2 d \tau^2 = ( 1 - \frac{r_s r}{\rho^2} ) c^2 dt^2 - \frac{\rho^2}{\Delta} dr^2 - \rho^2 d \theta^2 -
                  (r^2 + \alpha^2 + \frac{r_s  r \alpha^2}{\rho^2}  \sin^2 \theta)  \sin^2 \theta d \phi^2 +
                  \frac{2 r_s  r \alpha \sin^2 \theta}{\rho^2} c dt d\phi
\end{equation}
where
\begin{equation}
r_s=\frac{2GM}{c^2}
\end{equation}
is the Schwarzschild radius which defines the event horizon for an uncharged blackhole with zero angular momentum.  In addition, the following quantities:
\begin{equation} \label{eqn:Kerrconstants}
\begin{split}
\alpha &  = \frac{J}{Mc} \\
\rho^2 & = r^2 + \alpha^2 \cos^2 \theta \\
\Delta & = r^2 - r_s r + \alpha^2
\end{split}
\end{equation}
provide insight regarding the behavior of the metric components. For instance, for a muon neutrino, appropriate parameter values correspond to $|J|=\hbar / 2$ and $M \approx 170 keV/c^2$.  Under these circumstances, $r_s= 4.48 \times 10^{-58}\; m$ and $\alpha= 5.81 \times 10^{-13}\; m$.

\subsection{Spacetime Compression} \label{sec:stcompression}
For a non-rotating body, using spherical coordinates, the $dr^2$
term in the Schwarzschild metric contains the factor $g_{rr}=-(1-
\frac{r_s}{r})^{-1}$.  Consider two concentric spheres in the
geometry defined by the Schwarzschild metric. Along the radial
direction, with time held constant, an observer at rest would
measure a distance of $(1- \frac{r_s}{r})^{-1/2} dr$. Hence, as the
radial distance decreases from $r=\infty$ to $r=r_s$, flat spacetime
becomes 'stretched' by the factor $(1- \frac{r_s}{r})^{-1/2}$.  For
the case of the Kerr metric, this effect appears to be quite
different due to the introduction of the angular momentum term
$\alpha$  in Equation (\ref{eqn:Kerrconstants}).  Under these
circumstances, the Kerr metric in Equation (\ref{eqn:Kerrmetric})
yields the term $g_{rr}=-\frac{\rho^2}{\Delta}$ (which reduces to
the Schwarzschild term when $\alpha=0$). For the Schwarzschild case
$|g_{rr}|>1$ as long as $r>r_s$.  However, this is not necessarily
the case for the Kerr metric. Observe, a spacetime compression can
take place whenever the condition $\frac{\rho^2}{\Delta} < 1$ is
satisfied. Let us investigate how such a condition for spacetime
compression might arise by considering Equation
(\ref{eqn:Kerrconstants})
\begin{equation} \label{eqn:compressionderivation}
\begin{split}
\rho^2 < \Delta &  \Rightarrow r^2 + \alpha^2 \cos^2 \theta < r^2 - r_s r + \alpha^2 \\
 &  \Rightarrow \alpha^2 (1 - \cos^2 \theta) > r_s r
\end{split}
\end{equation}
and, hence, all values of $r$ such that
\begin{equation} \label{eqn:compressioncondition}
 r < \frac{\alpha^2 \sin^2 \theta}{r_s}   \equiv r_C
\end{equation}
will lead to a factor in $g_{rr}$ such that spacetime is compressed.
Given the parameter values from Section \ref{sec:Kerr metric}, it is
clear that such a condition can be satisfied. In fact, when $0 < r <
r_C$, it follows that $0 < \frac{\rho^2}{\Delta} < 1$.  Clearly, for
some values of $\theta$, there will be no compression effect. For
instance, if $\theta=0$, then $r_C=0$. On the other hand, given a
neutrino ensemble, the effect should be observable for particles
where $\theta \rightarrow \pi/2$.

\begin{figure}[ht!]
\centering
\includegraphics[scale=.55]{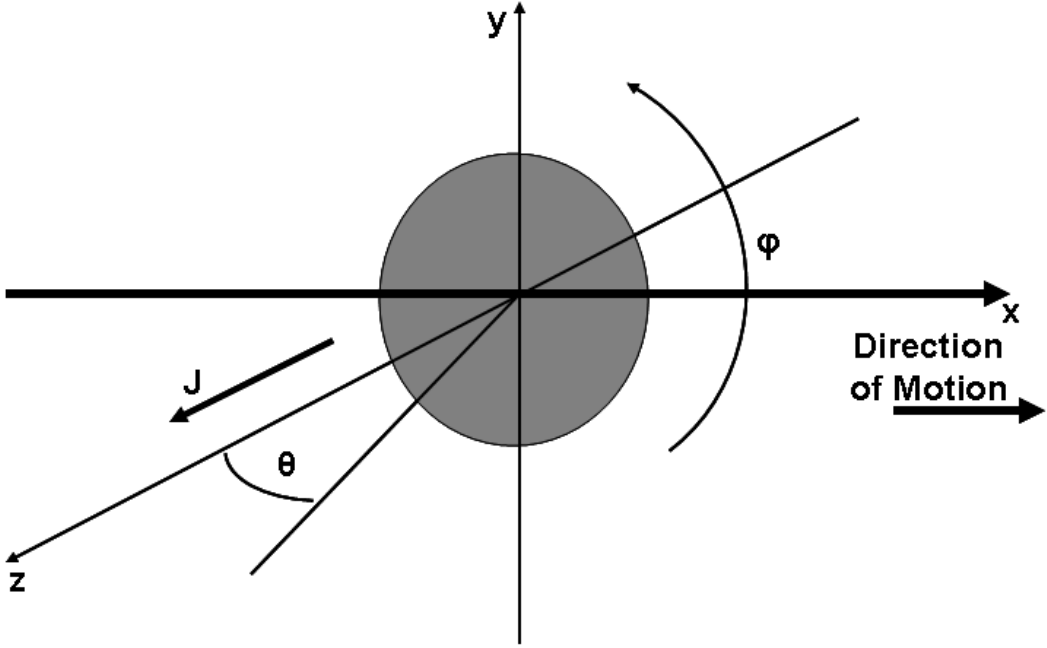} 
\caption{Particle in spherical coordinates where the spin vector is
aligned with the $z$-axis and direction of motion is along the
$x$-axis.} \label{fig:neutrinofig1}
\end{figure}

Rigorously verifying such an effect necessarily requires more
accurately describing the particle behavior at the quantum level. In
this presentation, we temporarily circumvent this more rigorous path
in order to present an approximate framework using the Kerr metric
and investigating the feasibility of such a hypothesis. Let us
consider the case of maximum compression where $\theta = \pi/2$.
Observe, in rectangular coordinates (e.g.$ \{ x,y,z \}$) this
corresponds to the spin vector $\vec{J}$ aligned with the $z$-axis
(see Figure \ref{fig:neutrinofig1}). Assuming the direction of
motion to be in the $x-y$ plane, and, without loss of generality,
along the $x$-axis, spacetime compression would take place along the
direction of motion. The obvious goal then is to relate the radial
compression factor to $\gamma_\nu$ in Equation
(\ref{eqn:gammacondition}).

Then, it must also follow that
\begin{equation}
\sqrt{\frac{\rho^2}{\Delta}} |_{\theta=\frac{\pi}{2}} = \sqrt{\frac{r^2 }{r^2 - r_s r + \alpha^2 }} = \gamma_\nu,
\end{equation}
or, in other words,
\begin{equation} \label{eqn:grrgammacondition}
\gamma_\nu = \sqrt{\frac{\rho^2}{\Delta}} |_{\theta=\frac{\pi}{2}} = \sqrt{\frac{1 }{1 + \frac{\alpha^2 - r_s r }{r^2}}} ,
\end{equation}
As in the previous section, assuming $\eta_\nu \approx .999995$, Equations (\ref{eqn:gammacondition}) and (\ref{eqn:grrgammacondition}) then tell us that
\begin{equation} \label{eqn:radialcondition}
\frac{\alpha^2 - r_s r }{r^2}  \approx 10^{-4}.
\end{equation}
Therefore, given the parameters listed in Section \ref{sec:Kerr metric}, we arrive at the following approximate result:
\begin{equation} \label{eqn:compression radius}
r^{\ast} \approx \sqrt{10^{4} \alpha^2} \approx 10^{-11} \;\; m
\end{equation}
where  $r^{\ast}$ represents the effective neutrino radius that generates the experimentally observed compression factor $\gamma_\nu$.

\section{Evaluation within the Context of the Standard Model} \label{sec:standardmodel}
We next consider particle categories consistent with those observed within the context of the standard model in order to understand why the observation of the compression effect has been limited to the neutrino.  As stated above, the main reason is most likely due to the fact that the neutrino is an uncharged lepton with an enormous mean free path length. Many particles in the standard model can immediately be eliminated based simply upon the short lifetimes. Particles with long lifetimes generally have mean free paths much shorter than the neutrino. In addition, charged particles will necessarily interact with the electromagnetic field.

\subsection{Uncharged Particles with Nonzero Spin}
In the case of particles having mass, the candidate metric to apply is the Kerr metric  \cite{oneill,steph} which is applicable to uncharged bodies with mass possessing nonzero angular momentum.

\begin{itemize}
\item \textbf{Neutrons}: The neutron might also be a candidate for observing the compression effect. If so, the implications regarding the formation of dark matter at the time of the big bang would have to be investigated (see Discussion section below). On the other hand, larger neutron mass and size might also lead to factors preventing it. If Equations (\ref{eqn:compressionderivation}) and (\ref{eqn:compressioncondition}) are applied to the neutron, $r* \approx 10^{-14}$ might prevent the effect given the size of the neutron. In addition, the quark structure of the neutron may be a factor in precluding the use of the Kerr metric as an approximation to quantum gravitational effects.
\item \textbf{Photons}: Photons are massless; hence, a spacetime compression effect would not be possible and the Kerr metric analysis would be applicable in this case.
\item \textbf{W Bosons}: The lifetime for these particles are much too short to observe such an effect.
\end{itemize}

\subsection{Charged Particles with Nonzero Spin}
For particles with charge, the electromagnetic force is assumed to overwhelm the gravitational effect.  The theory of general relativity allows for geodesics to be affected by charge. Under these circumstances, the Kerr-Newman metric is useful for describing massive charged bodies with nonzero angular momentum.

\begin{itemize}
\item \textbf{Electrons}:
Given the lifetime of the electron, it seems that the electron could be a candidate for the spacetime compression effect; however, it is likely that electromagnetic interactions could prevent the electron from establishing a long enough path to obseve the effect. Futhermore, as explained below, there may be general relativistic reasons due to frame dragging that inhibit the effect as well.
\item \textbf{Muons and Tauons}: The lifetime for these particles are much too short to observe such an effect.
\item \textbf{Z Boson}: The lifetime for this particle is much too short to observe such an effect.
\item \textbf{Protons}: In addition to mass, size and quark issues similar to that of the neutron mentioned above, electromagnetic interactions due to the proton charge may prevent observation of the effect.
\end{itemize}

\subsubsection{Kerr-Newman Metric}
It has been proposed that the Kerr-Newman metric is useful for describing the behavior of the electron \cite{burin08,burin11}.
The Kerr-Newman metric for a charged mass $M$ with charge $Q$ and angular momentum $J$ can be written using spherical coordinates $(r,\theta,\phi)$ as:
\begin{equation}
c^2 d \tau^2 = -(\frac{dr^2}{\Delta} + d \theta^2) \rho^2 +
               (c dt - \alpha \sin^2 \theta d \phi )^2 \frac{\Delta}{\rho^2} -
               ( (r^2 + \alpha^2) d \phi - \alpha c dt)^2 \frac{\sin^2 \theta}{\rho^2}
\end{equation}
where, once again,
\begin{equation}
r_s=\frac{2GM}{c^2}
\end{equation}
is the Schwarzschild radius. Furthermore, introducing the charge $Q$ leads to the metric parameters:
\begin{equation}
\begin{split}
\alpha &  = \frac{J}{Mc} \\
\rho^2 & = r^2 + \alpha^2 \cos^2 \theta \\
\Delta & = r^2 - r_s r + \alpha^2 + r_Q^2 \\
r_Q^2 & = \frac{Q^2 G}{4 \pi \epsilon_0 c^4}.
\end{split}
\end{equation}
Plugging in parameters leads to $r_Q^2 = 1.89 \times 10^{-72}$.

It will also be useful to express the Kerr-Newman metric as follows:
\begin{equation}
c^2 d \tau^2 = (g_{tt}-\frac{g^2_{t \phi}}{g_{\phi \phi}}) dt^2 + g_{rr}dr^2 + g_{\theta \theta}  d \theta^2 +
                  g_{\phi \phi} (d\phi  + \frac{g_{t \phi}}{g_{\phi \phi}} dt )^2
\end{equation}
where
\begin{equation}
\begin{split}
g_{rr} & = -\frac{\rho^2}{\Delta} \\
g_{\theta \theta} & = - \rho^2 \\
 g_{\phi \phi}  & = -(r^2 + \alpha^2 + \frac{(r_s  r - r_Q^2) \alpha^2}{\rho^2}  \sin^2 \theta)  \sin^2 \theta \\
 g_{t \phi} & = \frac{ (r_s  r - r_Q^2) \alpha \sin^2 \theta}{\rho^2} c  \\
 g_{tt} & =  ( 1 - \frac{(r_s  r - r_Q^2)}{\rho^2} ) c^2
 \end{split}
\end{equation}
As $Q \rightarrow 0$, $r_Q \rightarrow 0$ and the above equations reduce to those of the Kerr metric. Assuming $Q \neq 0$, Equations (\ref{eqn:compressionderivation}) and (\ref{eqn:compressioncondition}) can be rederived using $-\frac{\rho^2}{\Delta}$ for the Kerr-Newman metric. Under these circumstances, the compression condition becomes
\begin{equation} \label{eqn:KNcompressioncondition}
 r < \frac{(\alpha^2+r_Q^2) \sin^2 \theta}{r_s}   \equiv r_{QC}.
\end{equation}
Since the $r_Q^2$ term is small with respect to $\alpha^2$, one might arrive at a conclusion similar to that for the Kerr metric; however, we point out an issue we believe sheds further light on the compression effect.

\section{Frame Dragging} \label{sec:framedragging}
\begin{figure}[ht!]
\centering
\includegraphics[scale=.65]{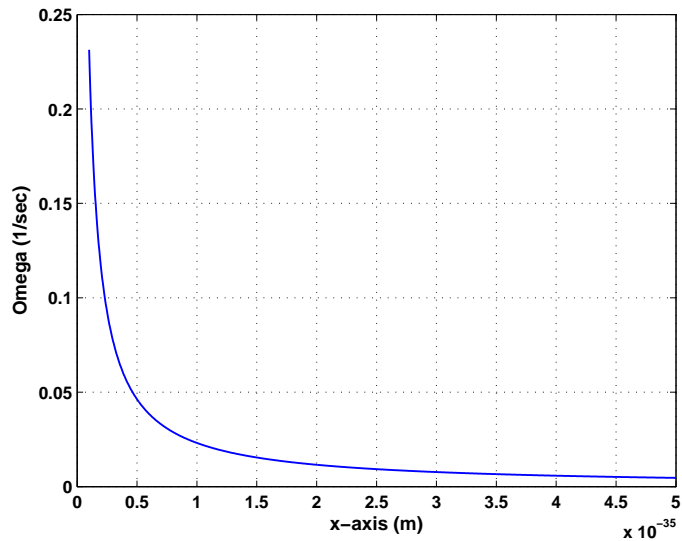}
\caption{Equation (\ref{eqn:Kerrframedragging}) plotted along $\theta=\pi/2$ direction of motion.}
\label{fig:neutrinofig2}
\end{figure}

\begin{figure}[ht!]
\centering
\includegraphics[scale=.65]{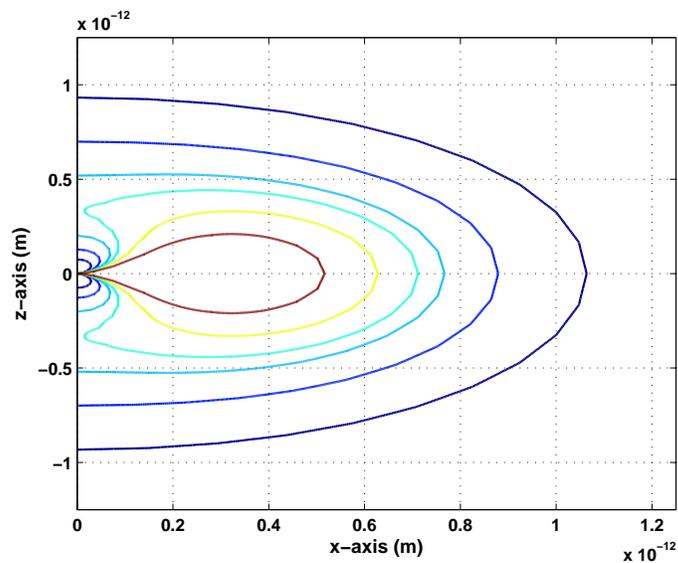}
\caption{Contour plot of values where $\Omega=$ constant in Equation
(\ref{eqn:Kerrframedragging}). Contours having the same color
correspond to equal values of $\Omega$.} \label{fig:neutrinofig3}
\end{figure}

Using the Kerr-Newman metric, frame dragging effects can be computed from the quantity:
\begin{equation} \label{eqn:KNframedragging}
\Omega = -\frac{g_{t \phi} }{g_{\phi \phi}}=\frac{(r_s r - r_Q^2)\alpha c}{\rho^2(r^2+\alpha^2)+(r_s r - r_Q^2) \alpha^2 \sin^2 \theta}
\end{equation}
When $Q=0$, this leads to $r_Q=0$ and Equation
(\ref{eqn:KNframedragging}) reduces to the frame dragging
relationship typically encountered using the Kerr metric:
\begin{equation} \label{eqn:Kerrframedragging}
\Omega =\frac{r_s r \alpha c}{\rho^2(r^2+\alpha^2)+r_s r  \alpha^2 \sin^2 \theta}
\end{equation}
As described in Section \ref{sec:stcompression}, with the spin
vector aligned along the $z$-axis, spacetime rotation takes place in
the $\phi$ direction (see Figure \ref{fig:neutrinofig1}).
Furthermore, along the direction for maximum compression outlined
above, the limit of Equation (\ref{eqn:Kerrframedragging}) as $r
\rightarrow 0$ along the direction of motion (where $\theta=\pi/2$)
is $\Omega \rightarrow c/ \alpha \approx 10^{20} s^{-1}$. Near the
Planck length, $r \approx 10^{-35} m$ and $\Omega \approx .025
s^{-1}$, then $\Omega$ begins rapidly approaching its limit (see
Figure \ref{fig:neutrinofig2}). The behavior of Equation
(\ref{eqn:Kerrframedragging}) is interesting in that $\Omega
\rightarrow 0$ as $r \rightarrow 0$ from values of $\theta \neq
\pi/2$. In addition, it is also true that values of $\Omega$ drop
off rapidly as $\theta$ moves away from $\pi/2$.  This behavior is
illustrated in Figure \ref{fig:neutrinofig3} which shows contours of
constant $\Omega$ in the $x-z$ plane near the value of $r^\ast$
derived in Equation (\ref{eqn:compression radius}). As the angle
diverges from $\theta = \pi/2$, Equation
(\ref{eqn:Kerrframedragging}) behaves like $r/k_1$ as $r \rightarrow
0$ and $k_2/r^3$ as $r \rightarrow \infty$  (where $k_1,k_2$ are
positive constants) with a single maximum between these two limits.
Hence, for any angle, there will exist two contours yielding the
same angular frequency. Near $\theta = \pi/2$, however, $\Omega$
behaves as in Figure \ref{fig:neutrinofig2} and values of $\Omega$
are substantially larger along the assumed direction of motion.
Hence, Kerr metric-based frame dragging appears to give results
consistent with the spacetime compression formulation presented
above when motion is along the $\theta=\pi/2$ direction.  We
therefore propose that strong frame dragging at scales near the
Planck length may serve as a potential mechanism for initiating the
observed spacetime compression.

On the other hand, for the
Kerr-Newman metric, an interesting effect arises that differs in
behavior from the Kerr metric. In this instance, we see that when
\begin{equation} \label{eqn:KNfdreversal}
r=\frac{r_Q^2}{r_s} =  \frac{1}{ 2 \pi \epsilon_0 c^2} \frac{Q^2}{M}
\end{equation}
is applied in Equation (\ref{eqn:KNframedragging}),
the direction of rotation reverses as $r < \frac{r_Q^2}{r_s}$ changes to $r >\frac{r_Q^2}{r_s} $. A similar conclusion has been
published in \cite{klein}.
Given the parameter values from Section \ref{sec:Kerr metric},
Equation (\ref{eqn:KNfdreversal}) implies that at
$r \approx 10^{-14} m >>$ electron radius, the direction of rotation
reverses. In addition, the $g_{tt}$ term in the Kerr-Newman
metric changes from a compression to an expansion. We hypothesize that this shift in the angular rotation
about the spin axis (as in the previous section, assumed to be along
the z-axis) due to electromagnetic interactions with spacetime may
inhibit the compression effect (which requires $r^\ast \approx
10^{-11}$ for the electron mass).

\section{Discussion} \label{sec:discussion}
The goal of this work has been to present a conceptual framework for
understanding the faster-than-light neutrinos in terms of a path
length compression due to general relativistic effects.  Given a
century of experimental precedent, the unwillingness to give up $c$
as a postulate of relativity is well-founded. Either the neutrino
experiments are wrong or some effect needs to be further understood.
Hence, we propose that general relativity at the quantum scale has
the capacity to resolve the experiments if the neutrino effect is
real. Under these circumstances, we view these results as an
opportunity to help understand spacetime at scales approaching the
Planck length. Specifically, we have proposed that strong frame
dragging at scales near the Planck length may serve as a potential
mechanism for initiating the observed spacetime compression.

One issue to be resolved is whether the neutrino should be modelled
as a black hole or a singularity. Such questions can only be resolved with a more in
depth formulation involving the Dirac equation. For the purposes of
this work, we only require the above framework in order to derive
approximate relationships. When the parameters described in
this work are applied to the Kerr metric, conclusions similar to
those described for the electron  \cite{burin08,burin11} can be
reached in that the neutrino could also be modelled as a
singularity. For instance, one might seek to describe the event horizon
by searching for values of $r$ such that $\Delta \rightarrow  0$ in Equation
(\ref{eqn:Kerrmetric}) such that
\begin{equation}
r=\frac{r_s \pm \sqrt{r_s^2-4\alpha^2}}{2}
\end{equation}
however, the parameters applied in this analysis would preclude the possibility of
$\Delta \rightarrow  0$. On the other hand, a change from timelike to spacelike Kerr metric behavior
can occur if
\begin{equation}
r=\frac{r_s \pm \sqrt{r_s^2-4\alpha^2 \cos^2 \theta}}{2}
\end{equation}
Values where $\theta \rightarrow \pi/2$ would allow for this possibility for values where
\begin{equation}
\cos \theta < \frac{r_s}{2 \alpha}.
\end{equation}
This of course would place limits on the physical
geometry of the particle such that $|\theta - \pi/2|$ is extremely small, possibly pointing to singular behavior \cite{burin08,burin11}.  Observe that
the geometric constraint $\theta \rightarrow \pi/2$ leading to singular behavior is consistent with the
conditions for maximum compression presented in Section \ref{sec:stcompression} and Section \ref{sec:framedragging}.

Given the approximate description presented above, the door has been opened
for hypothesizing more accurate descriptions of
neutrino-spacetime interactions. For
instance, particle frame dragging effects near the Planck length
could lead to the formation of 'micro-wormholes' through which the
particle might travel. Since the Dirac equation does allow for
spacetime tunneling effects, it may be interesting to formulate the
field statistics of micro-wormholes. Hence, a quantum field
description of spacetime compression might be formulated by
addressing phase transitions where the spin frame dragging effect
literally 'tears' into spacetime forming the
micro-wormhole. Under these circumstances, rather than the
compression appearing to be continuous at the macroscopic level, the
effect would actually be statistical. In other words, similar to how
a flat rock skips across a water surface, the particle skips through
a series of micro-wormholes. In analogy to a sling-shot effect, the
particle could possibly gain energy between skips leading to
different path compression lengths.

The implications of the observations are manifold as they
open up the possibility of developing techniques for engineering
spacetime travel using neutrino-spacetime interactions. It may even be possible to
amplify the effect by the presence of a local magnetic field designed to align the spins of a neutrino ensemble.
Finally, if the spacetime path length compression effect is real, then the
missing matter (e.g. dark matter, dark energy) problem might be
resolvable by considering particles capable of compressing spacetime
along their direction of motion. Clearly, after the big bang, an
immense amount of gravitational neutrino mass (and possibly mass due
to other particles with long enough lifetimes) travelling faster
than photons would exist beyond the visible universe.

\section*{Ackowledgements}
The first author would like to acknowledge  Dr. Robert Pompi of Binghamton University for his tantalizing
comments regarding neutrinos traveling faster than the speed of light after Supernova 1987A observations
were reported.

\bibliographystyle{plain}
\bibliography{neutrinorefs}

\end{document}